\documentclass{article}
\usepackage{ijcai22}

\usepackage{times}
\usepackage{soul}
\usepackage{url}
\usepackage[hidelinks]{hyperref}
\usepackage[utf8]{inputenc}
\usepackage[small]{caption}
\usepackage{graphicx}
\usepackage{amsmath}
\usepackage{amsthm}
\usepackage{booktabs}
\usepackage{algorithm}
\usepackage{algorithmic}
\usepackage{xcolor}
\usepackage{soul}

\title{Enhancing Oceanic Variables Forecast in the Santos Channel by Estimating Model Error with Random Forests}

\author{
    Author Name
    \affiliations
    Affiliation
    \emails
    pcchair@ijcai-22.org
}

\author{
Felipe M. Moreno\footnote{Main and corresponding author.
All other authors made significant suggestions and the last five authors
oversaw the whole project and secured funding.}$^1$\and
Caio F. D. Netto$^1$
\and
Marcel R. de Barros$^1$\and
Jefferson F. Coelho$^1$\and \\
Lucas P. de Freitas$^1$\and
Marlon S. Mathias$^2$\and
Luiz A. Schiaveto Neto$^1$\and
Marcelo Dottori$^3$ \and \\
Fábio G. Cozman$^1$\and
Anna H. R. Costa$^1$\and 
Edson S. Gomi$^1$\and
Eduardo A. Tannuri$^1$
\affiliations
Center for Artificial Intelligence (C4AI) -- University of Sao Paulo, Brazil\\
$^1$Escola Politécnica -- University of Sao Paulo, Brazil\\
$^2$Instituto de Estudos Avançados -- University of Sao Paulo, Brazil \\
$^3$Instituto Oceanográfico -- University of Sao Paulo, Brazil\\
\emails
\{felipe.marino.moreno, caio.netto, marcel.barros, jfialho, lfreitasp2001, marlon.mathias, fgcozman, anna.reali, eduat, mdottori\}@usp.br
}
\begin{document}

\maketitle

\begin{abstract}
In this work we improve forecasting of Sea Surface Height (SSH) and current velocity (speed and direction)
in oceanic scenarios. We do so by resorting to Random Forests so as to predict the error of a numerical forecasting system developed for the Santos Channel in Brazil. We have used the Santos Operational Forecasting System (SOFS) and data collected in situ between the years of 2019 and 2021. In previous studies we have applied similar methods for current velocity in the channel entrance, in this work we expand the application to improve the SHH forecast and include four other stations in the channel. We have obtained an average reduction of 11.9\% in forecasting Root-Mean Square Error (RMSE) and 38.7\% in bias with our approach. We also obtained an increase of Agreement (IOA) in 10 of the 14 combinations of forecasted variables and stations. 
\end{abstract}

\section{Introduction}

Forecasting of metocean conditions in coastal regions and waterways is an essential task in planning coastal and navigation operations. Forecasts of current and sea surface height (SSH) of water bodies have traditionally been made through numerical models that rely on the solution of simplified Navier-Stokes equations. Those models have inherent errors due to simplifications and uncertainties in parameters and boundary conditions.

%Due to the chaotic nature of such systems small uncertainties in boundary conditions and model parameters can lead to forecasts diverging from the ground truth.

An alternative to numerical models is to use machine learning (ML) to infer patterns in previous data measured in the region of interest and thus provide a forecast based only on the interpolation and extrapolation of those patterns observed in the past. However, since ML models only rely on correlations between data, thus ignoring the underlying physics of the problem, they fail if there is a change in the distribution of the data.

% Another possible line of work proposes the combination of machine learning with physics-based models (Physics Informed Machine Learning) that aims to extract the advantages of both the Machine Learning and Physics based model, namely the pattern recognition power and the generalization power in unseen scenarios.
A recent and promising line of work consists of combining ML with physics-based models --- often referred to as Physics-Informed Machine Learning (PIML). Such an approach aims to take advantage of both the power of pattern recognition given by ML approaches and the power of generalization in unseen scenarios given by the physics-based model.

% This work describes an implementation of Physics Informed Machine Learning where the ML is used to predict the error in water current velocity, direction and tides predicted by a physics based model. 
% %%Sugestão: This work expands on our previous PIML model (ref.) in which ML is used to predict the error in the speed and direction of water current and tides predicted by a physics-based model.
% %%%MAS precisa colocar mais claramente QUAL era a fraqueza do modelo anteriormente proposto e no que nossa proposta atual o alterou para melhorar a tal fraqueza. As frases seguintes, na minha opinião, não acrescentam nem valorizam muito o trabalho aqui proposto.
% By having a direct estimate of the model error it is possible to correct the model and improve forecasting accuracy. When compared to previous works in the region, this work increases the number of stations covered and adds correction for forecasting tides.

This work expands on our previous work \cite{9775449} where PIML was used to correct the error predicted by a numerical model of the speed of water current in a measuring station. Our main contribution here consists of inserting a correction for the direction of the water current and the sea surface height (SSH) predicted by the numerical model into the PIML model. In addition, we expand  the corrections to other measurement stations in the Santos-São Vicente-Bertioga Estuarine System region on the Brazilian coast. By producing a direct estimate of the numerical model error, one can correct the model and improve the prediction accuracy.

%This paper is structured as follows.
Section \ref{sec:related-work} introduces works that also used ML models to correct predictions made by numerical models.
Section \ref{sec:proposal} explains in detail our proposal, while 
Section \ref{sec:experiments} describes the experimental
setup and the experiments conducted. 
Section \ref{sec:results} presents the results of the experiments performed and some discussion.
Finally, Section \ref{sec:conclusions} presents our conclusions and highlights future work.

\section{Related Work}
\label{sec:related-work} 
% The combination of machine learning with physics-based models is a vast field of study, review papers such as the ones made by Williard et Al. and Kashinath et Al. that compile a myriad of applications already published that incorporates ML with physics based models with different objectives. Some works related to this one are briefly discussed below.

PIML is a relatively recent but already vast field of study; review articles such as those by \cite{Willard2020IntegratingSurvey} and \cite{Kashinath2021Physics-informedModelling} compile a myriad of applications that already embed ML with physics-based models for different purposes.

% Some applications of forecasting take benefit of simpler ML algorithms such as Random Forests to predict forecasting error, such the work of \citeauthor{Xu2015Data-drivenModel} where Quantile Random Forests (QRF) and Support Vector Machine (SVM) are used to predict groundwater base flow, those methods were used to provide a confidence interval of the prediction of the model error. \cite{Xu2015Data-drivenModel}
Some forecasting applications benefit from using traditional ML techniques to improve the results of physics-based models. For example, \cite{Xu2015Data-drivenModel} use Random Forests (RF) and Support Vector Machine (SVM) to improve the predictive accuracy of groundwater flow numerical models and provide more robust prediction intervals. We now list a few other proposals in the literature that are most relevant to our purposes.

% Random Forests have been used to improve temperature forecasts in Alpine regions in northern Italy in the work of \citeauthor{Eccel2007PredictionModels}, where other ML algorithms were used to predict forecast model errors, such as Artificial Neural Networks (ANN) and Multi-Linear Model. In this work all models performed similarly. 
\cite{Eccel2007PredictionModels} perform a comparison of linear and nonlinear ML models as methods for post-processing the direct outputs of numerical weather forecast models to reduce the biases introduced by a coarse horizontal resolution. The system was used to predict minimum temperatures in a   region of the Italian Alps. Artificial Neural Networks (ANN), RF, and a Multi-Linear Model were evaluated, showing similar performance.

% Random Forests and SVM were also used to improve maximum and minimum temperature forecasts in the city of Seoul (South Korea) in the work of \citeauthor{Cho2020ComparativeAreas}, where those models were compared to Artificial Neural Networks (ANN) and a Multi-Model Ensemble of the three models. In this work the ensemble model worked best, followed by the ANN and SVM. 
\cite{Cho2020ComparativeAreas} evaluate the use of RF, SVM, ANN, and a multi-model ensemble to correct a numerical weather prediction model that outputs next-day maximum and minimum air temperatures in Seoul, South Korea. Hence ML is used to mitigate the systematic bias in air temperature forecasting caused by a coarse grid resolution and lack of parameterizations of the numerical model. The study showed that the multi-model ensemble had better generalization performance than the three single ML models.

% Many other authors have also applied variants of ANN to correct physics-based models. In the work of \citeauthor{Bonavita2020MachineCorrection} a Multi-Layer Perceptron ANN with three layers is used to correct models of temperature and pressure. 
As shown in the literature, the physical model error is one of the main obstacles to improving the accuracy and reliability in numerical weather and climate prediction. \cite{Bonavita2020MachineCorrection} use  Multi-Layer Perceptron ANN with three layers to model error estimation and correction in the numerical model temperature and pressure prediction.
% \cite{Vashani2010ComparativeIran} compares ANN with other models such as Kalman Filter and Moving Average as post processors for temperature over Iran, where it was shown that ANN gives the best results. 
\cite{Vashani2010ComparativeIran} make a comparative evaluation of different ML post-processing models for numerical prediction of temperature forecast over Iranian territory, concluding that ANN provides the best results.
% The work of \citeauthor{Isaksson2018ReductionNetworks} uses Convolutional Neural Networks (CNN) to reduce forecasting errors of temperature in the Scandinavian Peninsula, the CNN receives a grid of forecasted values of temperature and other environmental parameters, and produces forecasts with smaller errors than the ones obtained with Kalman Filter post-processor. 
Another type of ANN, Convolutional Neural Networks (CNNs), have been used to reduce temperature forecasting errors in the Scandinavian Peninsula \cite{Isaksson2018ReductionNetworks}. In that work a CNN receives a grid of forecasted values of temperature and other environmental parameters, and produces forecasts with smaller errors than those obtained with a Kalman filter post-processor. 
% In the work of \citeauthor{Chapman2019ImprovingLearning}, CNNs have been employed to learn and correct forecasts of vapor transport in North America. CNNs have also been used to predict forecasting uncertainy and spread for wind and 500hPa geopotentital heigh, as shown in the work of \citeauthor{Scher2018PredictingLearning}. \cite{Bonavita2020MachineCorrection,Vashani2010ComparativeIran,Isaksson2018ReductionNetworks,Chapman2019ImprovingLearning,Scher2018PredictingLearning}
CNNs have also been employed by \cite{Chapman2019ImprovingLearning} to learn and correct North American atmospheric river forecasts, and by \cite{Scher2018PredictingLearning} to predict uncertainty in weather forecast.

% Another work has applied QRF to predict model error in the Santos Channel, in the work of \citeauthor{Moreno2022} the forecast of surface current velocity in the entrance of the channel has been corrected. 
Finally, our previous work also applied RF in order to learn the error model of the forecast of sea surface current speed made by a numerical model for the entrance of the Santos Channel, in Brazil \cite{9775449}. In that work we only corrected the current speed in one station, while in the
present paper  we improve the PIML model by expanding its application to other stations and other two variables types, current direction and SSH.

\section{A Proposed SOFS + RF Architecture}
\label{sec:proposal}

\begin{figure}[t]
\centerline{\includegraphics[scale=0.175]{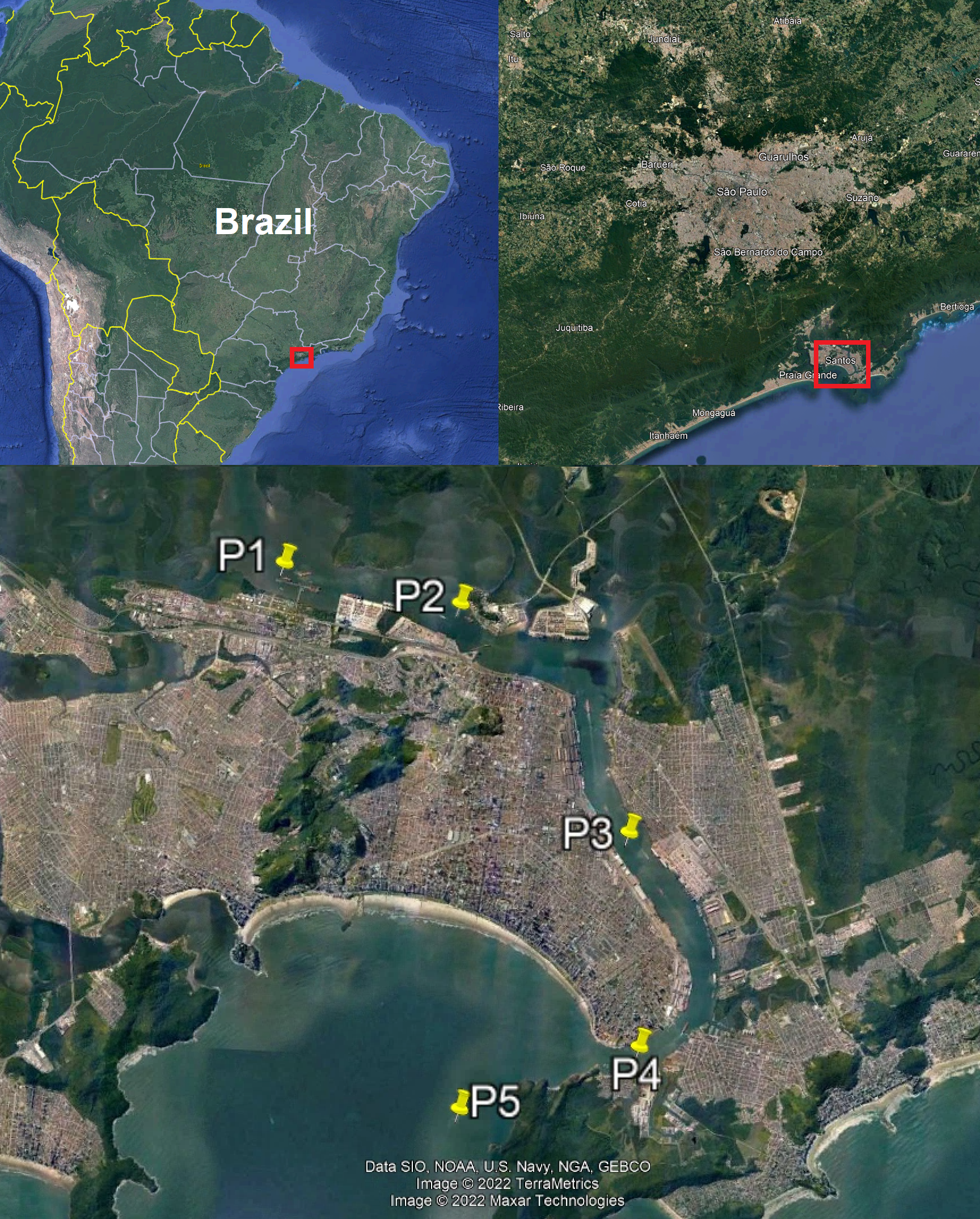}}
\caption{The top images are panoramic views of the Brazilian coast, to locate the Santos Channel. The bottom image shows the Santos Channel, and the five markers indicate the location of measuring stations kept by the Port Authority. Map Source: Google Maps.} % \vspace{10pt}
\label{fig:map}
\end{figure}

We propose a PIML approach to improve the prediction made by the 
\textit{Santos Operational Forecasting System} (SOFS) \cite{Costa2020AnBrazil}
in the Santos Channel, a key sea region in Brazil. 
%The objective of this project is to improve the forecasting of currents and tides in t
The Santos Channel (see Figure \ref{fig:map}) provides access to the Santos Port Complex, the largest port in Latin America with an yearly handling of about 145 million tons of cargo. The channel contains 5 measurement stations maintained by the Marine Pilots, where current and SSH measurements are captured. In addition, there are weather sensors in the region.
A numerical model based on physics is implemented within SOFS so as to forecast relevant quantities.  

Physics-informed techniques in machine learning can incorporate the physics of a domain of interest in different forms, as described for instance by \cite{Willard2020IntegratingSurvey}. Some approaches incorporate  the physics directly into the architecture used to learn various quantities, for example by taking equations to guide the loss function of an ANN during training. Other approaches are inspired by the physical problem so as to guide the design of the architecture, for example by providing the same boundary conditions used for a numerical model solver as inputs of the ML. Another approach is to use a stand alone numerical model  and use the ML algorithm to correct its output by either estimating the model error or pondering it with other runs made with perturbed inputs.

% We propose a PIML architecture to improve the forecasting in the region that uses Random Forests to predict the error of the SOFS model. 
% The ML based post processor predicts the physics based model error for current absolute velocity, current direction and surface elevation. For each target variable one Random Forest was trained using the prediction of the physics based model and sensor data as inputs. 
Because we have a numerical model already in use for the variables we are interested in, we adopted the third approach by developing an architecture that uses Random Forests (RFs) as an ML model to estimate the SOFS error, which is then corrected in a post-processing phase. 
RF was chosen for its simplicity, effectiveness and efficiency, also demonstrated in previous work \cite{Eccel2007PredictionModels}. 
However, it is worth noting that any other ML model or even an ensemble of ML models could be used for this function, as for instance argued by \cite{Cho2020ComparativeAreas}.

\begin{figure*}[!h]
\centerline{\includegraphics[scale=0.38]{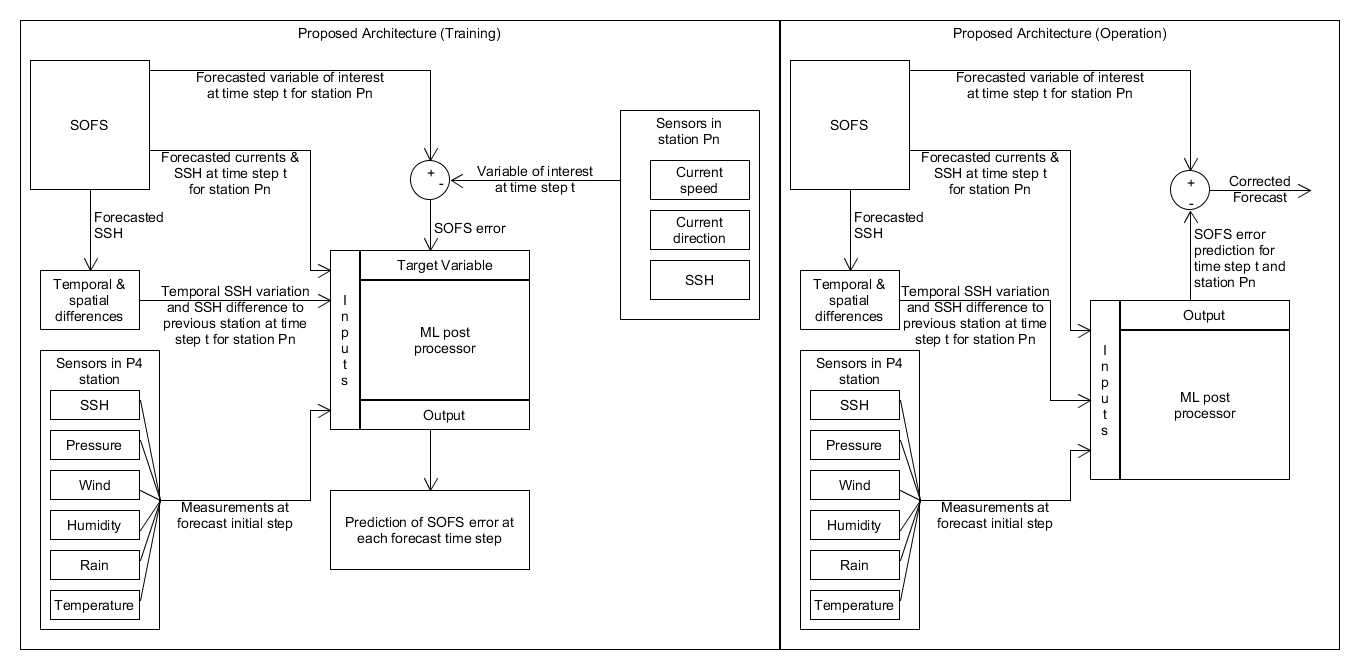}}
\caption{Our proposed Architecture. The training phase is illustrated in the left, where the ML model learns to estimate the SOFS error from the values of its input variables. The operation phase is illustrated in the right, which receives the input variables together with the SOFS estimate and provides the corrected estimate value at the output. There is an ML model trained specifically for each target variable and $P_n$ sensor station.}
\label{fig:arch}
\end{figure*}

The target variables estimated by our architecture are the current speed and direction, and SSH at each time step of the SOFS forecast.
A distinct RF is trained for each pair of target variable and measurement station $P_n$, with $n \in \{1, 2, 3, 4, 5\}$ (see Figure \ref{fig:map}), using the numerical model predictions and sensor data. % obtained in station P4 as inputs. 
Once the ML model is trained, the RFs are then used to correct the SOFS prediction error in the operation phase. 

An overview of the architecture is depicted in Figure \ref{fig:arch}.
The data used by the PIML architecture comes from the SOFS model and sensors. The SOFS model, the collected data, and the training and operating phases of the PIML architecture are described in the following.

\subsection{The Santos Operational Forecasting System}
\label{sec:SOFS}
The numerical model used for this work is the SOFS, a forecasting system based on the Princeton Ocean Model\footnote{http://www.ccpo.odu.edu/POMWEB/} named POM-rain module that provides forecasts of currents, SSH, salinity and temperature up to three days ahead for the Santos-Sao Vicente-Bertioga Estuarine System.
% The SOFS produces one forecast every day with an horizon up to 72 hours ahead, but for this work we will consider forecasts made with a horizon of 21 hours in time steps of 3 hours, totaling 8 steps in each day.

The SOFS model is based on the Navier-Stokes equation, considering Boussinesq approximation, hydrostatic pressure and an incompressible fluid. The system deals with the following equations:

\begin{equation} \label{eq:ns}
    \begin{cases}
    -\dfrac{\delta \eta}{\delta t} = \int_0^H \dfrac{\delta u}{\delta x} \delta z + \int_0^H \dfrac{\delta v}{\delta y} \delta z,   \\
        \dfrac{\delta u}{\delta t} + \vec{V} \cdot \nabla u -fv = -\dfrac{1}{\rho_0} \dfrac{\delta p}{\delta x} + \dfrac{1}{\rho_0}\nabla \tau_x, \\
        \dfrac{\delta v}{\delta t} +  \vec{V} \cdot \nabla v + fu = -\dfrac{1}{\rho_0} \dfrac{\delta p}{\delta y} + \dfrac{1}{\rho_0} \nabla \tau_y, \\
        \dfrac{\delta w}{\delta t} +  \vec{V} \cdot \nabla w = -\dfrac{1}{\rho_0} \dfrac{\delta p}{\delta y} + \dfrac{1}{\rho_0} \nabla \tau_y -\dfrac{\rho}{\rho_0 g}.
    \end{cases}
\end{equation}
where  $\eta$ is the SSH and $\vec{V} = [u,v,w]$ are water velocities in a Cartesian coordinate system where $x$ and $y$ axis are horizontal, and the $z$ axis is vertical. Other parameters of this equation are the total water column depth $H$, Coriolis acceleration $f$, water density $\rho$, water reference density $\rho_0$, gravitational acceleration $g$, pressure $p$ and stresses $\tau_i = [\tau_{ix},\tau_{iy},\tau_{iz}]$ in the direction $i$ due to both shear (such as viscosity and wind stress) and Reynolds stress. 

The first equation is the continuity equation, assuming water incompressibility; it indicates that the variation in water elevation at a given point is equal to the difference between the volume of water that enters and exits the water column at that point. 
The three remaining equations are the Navier-Stokes momentum balance in three directions, considering advection and Coriolis accelerations in the left side of the equation and pressure gradient, stresses and buoyancy forces in the $z$ direction in the right side.

\begin{figure}[!h]
\centerline{\includegraphics[scale=0.3]{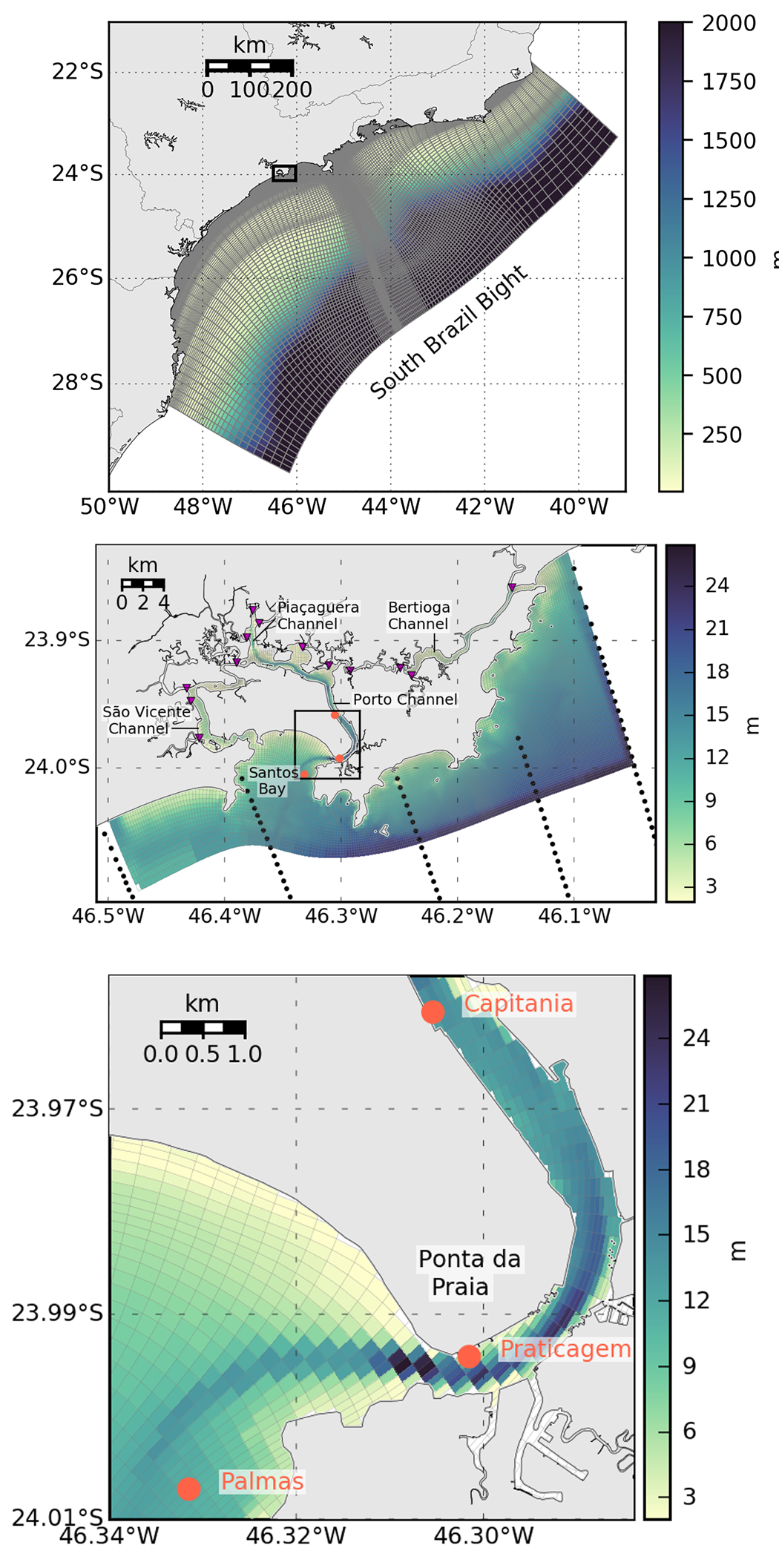}}
\caption{Top: Coarser grid of the SOFS encompassing the South Brazilian Bight with indication of the localization of the nested grid. Middle: nested fine grid for the Santos-Sao Vicente-Bertioga Estuarine system. Bottom: detail of the grid in the entrance of the Santos Channel, the points shown as Palmas, Praticagem and Capitania are respectively stations P5, P4 and P3 in Figure \ref{fig:map}. Source: \protect\cite{Costa2020AnBrazil}.}
\label{fig:sofs}
\end{figure}

SOFS works with two grids, one larger encompassing the coastal region from Southeast Brazil, and a nested grid of finer resolution, encompassing  the Santos-São Vicente-Bertioga Estuarine System, as shown in Figure \ref{fig:sofs}.

For the coarser grid, this model incorporates atmospheric boundary conditions from the Center for Weather Forecasts and Climate Studies (CPTEC, Portuguese acronym), currents in the open boundary is obtained from the Copernicus Marine Environment Monitoring Service Mercator (CMEMS), and tides in the boundary were obtained by astronomical components for the region. Boundary conditions for the nested grid are obtained from the coarser grid. It uses a three-dimensional grid with Sigma vertical coordinates and Arakawa C-grid for horizontal coordinates.

The SOFS model is also split into two modes, an external mode where the 2D equations are solved considering mean values for the entire water column and that is solved using a faster time-step of 0.8s, and an internal mode where the 3D equation system is solved in a time-step of 4 seconds.

The historical data available for this system is composed of daily forecast events, each one consisting of 8 forecasting steps with a time-step of 3 hours. The initial forecast step is around midnight GMT  and the last one is around 21:00 GMT. The forecast is available for all grid points of the system, as shown in Figure \ref{fig:sofs}, but for this work we selected only the points closest to the measuring stations as shown in Figure \ref{fig:map}. The selected period is between January $1^{st}$, 2019 and March $19^{th}$, 2021.

\subsection{In Situ Measured Data}
\label{sec:measured-data}
The measured current velocities and SSH used in this project were acquired from Sontek SL Acoustic Doppler Current Profilers (ADCP) installed in the channel; other measurements are obtained from a weather station installed in station P4 (see Figure \ref{fig:map}). 
The measurements available are 5 minute averages of surface current speed and direction for all the 5 stations and 5 minutes average, minimum and maximum values of wind, temperature, rainfall and relative humidity as well as average atmospheric pressure for the P4 station.

Data collection in the channel is carried out by the Santos Port Authority. Over time, sensors have been installed in the region to increase spatial coverage and the variety of data available. For this work, we selected the same time window used for SOFS, covering between January $1^{st}$, 2019 and March $19^{th}$, 2021, except for SSH at stations P2, P3, P5, where the measurement started December $1^{st}$, 2019, and for SSH at station P1 where no measurement is available. Data coverage for the selected time range depends on the data type and station. For example, the weather station at P4 has about 4.8\% missing data for wind, temperature, and rain, and 1.5\% for SSH in the selected time window.

\subsection{PIML training phase}
\label{sec:training}

\begin{table*}[htbp]
\begin{center}
\begin{tabular}{rrr|rrrrrrrrr|r}\hline
\multicolumn{3}{c|}{ \textbf{Measured Data} } & \multicolumn{9}{|c|}{ \textbf{SOFS Forecast Data} } & \multicolumn{1}{|c}{\textbf{Target Variables}}\\ \hline
 &  &  & \multicolumn{4}{|c|}{Forecast step} &  &  &  & SSH & SSH &   \\
Wind & $\hdots$ & Precip- & \multicolumn{4}{|c|}{(One-Hot} & Curr. & Curr. & SSH & temporal & spatial &  Current speed error \\ 
Spd. & & tation & \multicolumn{4}{|c|}{encoded)} & Spd. & Dir. & & difference & difference &  \\ \cline{4-7}
(m/s) & & (mm) & 0 & 1 & ... & \multicolumn{1}{r|}{7} & (m/s) & (\textdegree) & (m) & (m)  & (m) & (m/s)   \\ \hline
3.67 & ... & 0 & 1 & 0 & ... & \multicolumn{1}{r|}{0} & 0.35 & 92.3 & 2.11 & 0.40 & 0.06 & -0.03 \\
3.67 & ... & 0 & 0 & 1 & ... & \multicolumn{1}{r|}{0} & 0.24 & 85.6 & 1.71 & 0.56 & 0.07 & -0.06 \\
3.67 & ... & 0 & 0 & 0 & ... & \multicolumn{1}{r|}{0} & 0.13 & 248.6 & 1.15 & 0.12 & 0.24 & 0.08 \\
$\vdots$ & $\vdots$ & $\vdots$&$\vdots$ & $\vdots$& $\vdots$&\multicolumn{1}{r|}{$\vdots$} & $\vdots$& $\vdots$& $\vdots$& $\vdots$& $\vdots$& $\vdots$\\
\hline
\end{tabular}
\end{center}
\caption{Example of the training dataset structure for current speed as the target variable. Some measured data columns were omitted for brevity.} \label{tab:all}
\end{table*} 

In the training phase of each ML model, following the supervised paradigm, a training dataset is used consisting of pairs 
$\langle input \  variables\ , \ desired \  value \ of \  target \  variable \rangle$.

The measured target variable --- either current speed, current direction or SSH --- for station $Pn$, with $n \in \{1,2,3,4,5\}$, in time step $t$, with $t \in \{0,1, 2,...,7\}$, is given by $target_{P_n, t}$.
The ML model uses as input variables for the training phase the weather data observed at station P4 at the beginning of the forecast, and the SOFS forecast error for the respective target variable, station, and respective forecast step, $SOFSerror_{target , P_n, t}$.
The SOFS forecast error for the respective target variable, station, and respective forecast step, $SOFSerror_{target, P_n, t}$ is given by the difference between the value predicted by SOFS for the target variable, $SOFS_{target, P_n, t}$ and the measured value of the same target variable at the respective station, $target_{P_n, t}$.
For example, for the SSH target variable and station P5, comes:
\begin{equation*} 
SOFSerror_{SSH, P_5, t} = SOFS_{SSH, P_5, t} - SSH_{P_5, t}.
\end{equation*} 

Once the training database is built, given by pairs of the type
\begin{multline*}
    \langle \{weather-data_{t=0}, SOFS-data_{P_n, t}\\,target_{P_n, t} \}, SOFSerror_{target, P_n, t} \rangle,
\end{multline*}
the ML model is trained until a stopping condition is met.

The ML model trained from the PIML model can then be used in the operation phase.

\subsection{PIML operation phase}
\label{sec:operation}
In the operating phase, the ML model receives as input the same variables used as input in the training phase, and provides as output the SOFS error estimate for the respective target variable, measurement station and prediction time step, $SOFSerror_{target, P_n, t}$.

However, now the ML model output is subtracted from the prediction made by SOFS in the same step, for the same station and target variable, $SOFS_{target, P_n, t}$, resulting in a corrected prediction for the target variable as the output of the PIML system, $target_{P_n, t}$:
\begin{equation*}
   target_{P_n, t} = SOFS_{target, P_n, t} - SOFSerror_{target, P_n, t}.
\end{equation*}

\section{Experimental Setup}
\label{sec:experiments}

By merging both the SOFS and measured data it is possible to assemble a single training dataset for each combination of station $P_n$ and target variable to train the ML model. 
An example of the dataset structure is shown in Table \ref{tab:all}. 
Each row of the training dataset contains one forecasting step of the SOFS model for SSH and currents, the difference in forecasted SSH between the current and next SOFS step, the difference in forecasted SSH between the current and previous station in the channel, the forecast step in One-Hot Encoding, the measured values of wind, temperature, pressure, precipitation, relative humidity, and SSH in station P4 for the initial forecast step of SOFS, and the target variables. We decided to add the temporal and spatial SSH differences due to the SSH importance  in driving the currents in the channel.

To obtain the target variables, as explained in Section \ref{sec:training}, we subtract the measured values from the values predicted by the SOFS model, taking into account the closest measurement taken in time to the respective SOFS forecast step, within a maximum acceptable difference of 30 minutes. If there was no measurement available within 30 minutes of the forecast step time, we discarded the respective full day.
Other data treatments   would be possible, but our choice led to the desired analysis.

The direction error is expressed in the range between -180\textdegree\ and 180\textdegree, taking the angle wrapping around 0\textdegree (0\textdegree=360\textdegree).

The measured data in each row is the latest measurement available right before the first forecast step of the daily forecast event. If there is no measurement up to 30 minutes before the first forecast step, we discard the entire day. The measurement data that is used as input is composed of all variables measured in the weather station and the SSH variable at the station P4.

The ML model selected to predict the SOFS error is a Random Forest Regressor (RF), available in the Python Package Sklearn, due to the simplicity of tuning its hyperparameters and its characteristic of averaging the target variable values seen in the training dataset, avoiding predicted errors above what has been seen in the past. 
In a previous work \cite{9775449}, we used Quantile Regression Forests. However, 
%due to the deprecation of the Python Scikit-garden package, and the 
we fond in that previous work that the obtained uncertainty range was too wide to be useful, 
so we have decided to move to the RF model.

We ran a 5-fold cross validation using  random search to find the best hyperparameters of the RF model.
The random search approach was performed with 50 random samples at the intervals shown in the table \ref{tab:hyper}. 
We used the inital 80\% of the dataset for the cross-validation and model training, and reserved the remaining 20\% for testing. 

The hyperparameter optimization was carried on an Intel i7-11800h, taking on average 50 minutes to cross-validate the model for 50 points in the hyperparameter space.

\begin{table*}[htb]
\begin{center}
\begin{tabular}{lrrrrrr}
\hline
\multicolumn{1}{c|}{} & \multicolumn{3}{|c|}{ \textbf{SOFS IOA}} & \multicolumn{3}{|c}{ \textbf{SOFS+RF IOA}} \\ \cline{2-7}
\textbf{Station}& \multicolumn{2}{|c|}{ {Current}} &  & \multicolumn{2}{|c|}{ {Current}} &  \\ \cline{2-3} \cline{5-6}
\multicolumn{1}{c|}{}& \multicolumn{1}{|r|}{Speed} & \multicolumn{1}{r|}{Direction} &  SSH & \multicolumn{1}{|r|}{Speed} & \multicolumn{1}{r|}{Direction} & SSH \\ \hline
P1 & 0.409 ±0.151 & 0.334 ±0.175 & -            & 0.370 ±0.133 & 0.581 ±0.170 & - \\ 
P2 & 0.528 ±0.193 & 0.636 ±0.281 & 0.922 ±0.113 & 0.516 ±0.195 & 0.662 ±0.288 & 0.933 ±0.115\\ 
P3 & 0.563 ±0.203 & 0.688 ±0.159 & 0.880 ±0.141 & 0.556 ±0.200 & 0.671 ±0.162 & 0.886 ±0.137\\ 
P4 & 0.591 ±0.231 & 0.754 ±0.185 & 0.945 ±0.051 & 0.622 ±0.227 & 0.783 ±0.181 & 0.958 ±0.047\\ 
P5 & 0.483 ±0.193 & 0.585 ±0.221 & 0.944 ±0.083 & 0.553 ±0.182 & 0.668 ±0.206 & 0.953 ±0.076\\ \hline
\end{tabular}
\end{center}
\caption{Index of Agreement obtained with the combination of SOFS and Random Forests, compared with results obtained with the SOFS model alone.}\label{tab:res}
\end{table*}

\begin{table} [htb]
\begin{center}
\begin{tabular}{lrr}
\hline
\textbf{Hyperparameter} & \textbf{Interval} & \textbf{Steps}\\ \hline
Number of features& 3 - 15 &  2\\
Max. tree depth& 5 - 60 & 5 \\
Min. samples for split& 2 - 42  & 4 \\
Min. samples in leaf& 1 - 25 & 4 \\
Number of trees& 50 - 1500 & 50 \\
Loss Metric & MAE, MSE & - \\ \hline
\end{tabular}
\end{center}
\caption{Hyperparameters types and ranges tuned for the post processor. Random search was made inside the interval shown in this table. MAE and MSE stand for "Mean Absolute Error" and "Mean Squared Error", respectively.} \label{tab:hyper}
\end{table}

The random search selected the hyperparameters that maximized the Index of Agreement (IOA). 
The IOA is given by 
\begin{equation} \label{eq:ioa}
    IOA = 1 - \frac{\sum^n_{i=1}(O_i-P_i)^2}{\sum^n_{i=1} (|P_i-\overline{O}| + |O_i - \overline{O}|)^2},
\end{equation}
for a sequence of $n$ observations $O_i \in \{O_1, O_2,..., O_n\}$ with mean $\overline{O}$, and predicted values $P_i \in \{P_1, P_2,..., P_n\}$.

The training dataset was then used to train the RF model with the best hyperparameters found in the Cross-Validation.
The model was then evaluated in the test dataset, in the operation phase desribed in Section \ref{sec:operation}. The results os the evaluations carried out are described in the following.

\section{Results}
\label{sec:results}
We used the Wilcoxon test to decide whether the distribution of values forecasted by the SOFS model and the PIML model that combines SOFS with the RF model (SOFS+RF) are different. 

The Wilcoxon signed-rank test is a non-parametric hypothesis test that is used to verify if two distributions are different in a statistically significant way. 
To measure the model performance we decided to use the Index of Agreement (IOA), a typical performance metric used to evaluate metocean forecasting models. 
The IOA metric was calculated for each forecast day in the test dataset. IOA average and standard deviation is shown in Table \ref{tab:res}.

The results show that there is an increase in the IOA when the PIML model is employed. 
The amount of the increase depends on the parameter of interest and the location of the station.
For some stations, the improvement in the IOA for the current direction was marginal or non-existent. 
The improvement obtained for the SSH variable was also small, but this may be due to the high IOA values obtained using the SOFS model, which already shows very good effectiveness, leaving only a small margin for improvement.
Current speed prediction using the PIML system, on the other hand, showed a more consistent increase in the IOA metric for all stations.

We also have calculated the Root-Mean Square Error (RMSE) for both the SOFS model and the PIML system.
RMSE decreases for all variables and stations when using the PIML model, except for SSH predictions in the P3 station. One possible explanation for this behavior is a distribution shift between the training and test data.

\begin{table}
\begin{center}
\begin{tabular}{lrrrrrr}
\hline
 & \multicolumn{3}{|c|}{ \textbf{SOFS}} & \multicolumn{3}{|c}{ \textbf{SOFS+RF}} \\ \cline{2-7}
\textbf{Station}& \multicolumn{2}{|c|}{ {Current}} &  & \multicolumn{2}{|c|}{ {Current}} &  \\ \cline{2-3} \cline{5-6}
& \multicolumn{1}{|r|}{Speed} & \multicolumn{1}{|r|}{Dir.} &  \multicolumn{1}{|r|}{SSH} & \multicolumn{1}{|r|}{Speed} & \multicolumn{1}{|r|}{Dir} & SSH \\
& \multicolumn{1}{|r|}{(m/s)} & \multicolumn{1}{|r|}{(\textdegree)} &  \multicolumn{1}{|r|}{(m)} & \multicolumn{1}{|r|}{(m/s)} & \multicolumn{1}{|r|}{(\textdegree)} & (m) \\ \hline
P1 & 0.089 & 117.9 & - & 0.071 & 91.5 & -\\ 
P2 & 0.126 & 82.1 & 0.137 & 0.109 & 75.6 & 0.116 \\ 
P3 & 0.187 & 79.3 & 0.189 & 0.166 & 65.4 & 0.196\\ 
P4 & 0.178 & 72.3 & 0.125 & 0.159 & 62.4 & 0.105 \\ 
P5 & 0.090 & 86.8 & 0.125 & 0.074 & 75.7 & 0.109\\ \hline
\end{tabular}

\end{center}
\caption{Comparison of Root-Mean Square Error for both  SOFS and SOFS+RF.} \label{tab:res}
\end{table}

We also have calculated the Wilcoxon test to verify whether the distribution of predictions obtained with SOFS+RF is different from the SOFS model. 
If we consider the typical threshold of $p\leq0.05$ to reject the null hypothesis, we can observe that in the majority of cases the distribution obtained with the SOFS+RF system is different from the one produced with the SOFS model. 
It can be claimed that the PIML system (SOFS+RF) produces statistically better results than the SOFS model.

\begin{table}[htb]
\begin{center}
\begin{tabular}{lrrr}
\hline
\multicolumn{1}{c}{} & \multicolumn{3}{|c}{ \textbf{Wilcoxon Test}} \\ \cline{2-4}
\textbf{Station}& \multicolumn{2}{|c|}{ {Current}} &   \\ \cline{2-3}
\multicolumn{1}{c|}{}& \multicolumn{1}{|r|}{Speed} & \multicolumn{1}{r|}{Direction} &  SSH \\ \hline
P1 & $<0.001$ & $0.491$ & -  \\ 
P2 & $<0.001$ & $<0.001$ & $<0.001$\\ 
P3 & $0.001$  & $<0.001$ & $<0.001$ \\ 
P4 & $<0.001$ & $<0.001$ & $<0.001$\\
P5 & $<0.001$ & $0.055$  & $<0.001$\\  \hline
\end{tabular}
\end{center}
\caption{P-values for the Wilcoxon test for the distribution of predictions of SOFS and SOFS+RF.}\label{tab:wilcoxon}
\end{table}

\begin{figure*}[t]
\centerline{\includegraphics[scale=0.25]{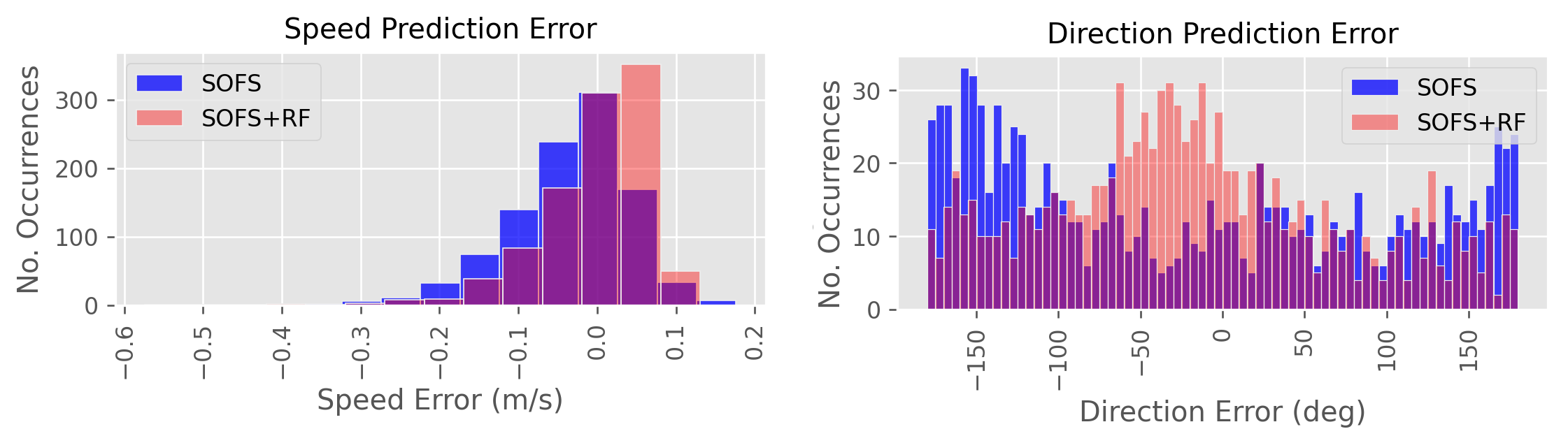}}
\caption{Forecast error distributions for   point P1.}
\label{fig:error_p1}
\end{figure*}

\begin{figure*}[t]
\centerline{\includegraphics[scale=0.20]{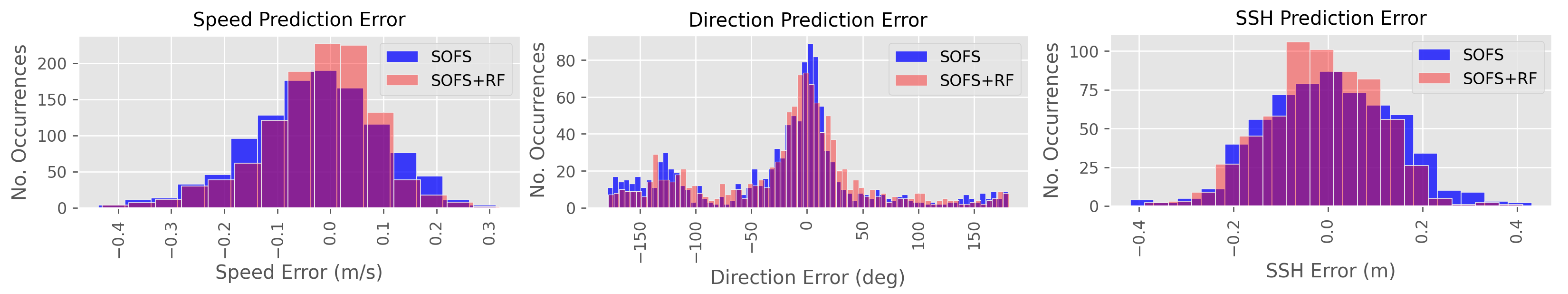}}
\caption{Forecast error distributions for   point P2.}
\label{fig:error_p2}
\end{figure*}

\begin{figure*}[!h]
\centerline{\includegraphics[scale=0.20]{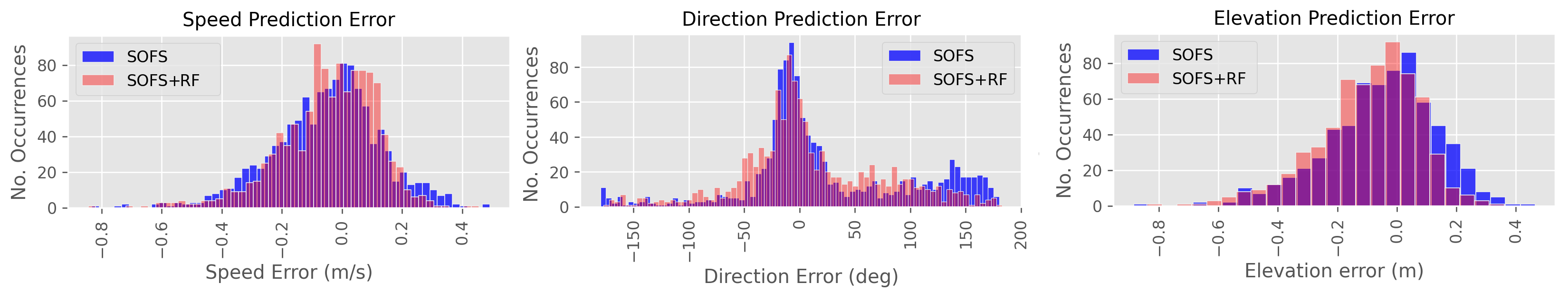}}
\caption{Forecast error distributions for   point P3.}
\label{fig:error_p3}
\end{figure*}

\begin{figure*}[!h]
\centerline{\includegraphics[scale=0.20]{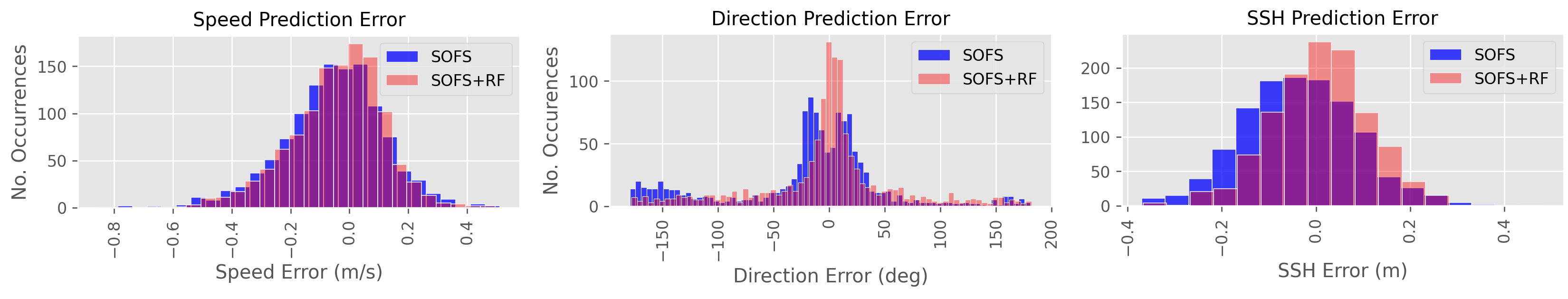}}
\caption{Forecast error distributions for   point P4.}
\label{fig:error_p4}
\end{figure*}

\begin{figure*}[!h]
\centerline{\includegraphics[scale=0.20]{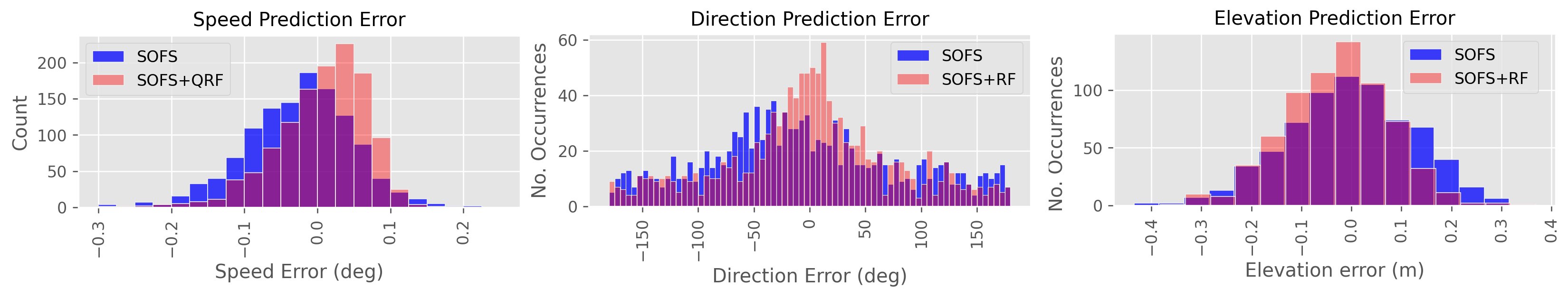}}
\caption{Forecast error distributions for   point P5.}
\label{fig:error_p5}
\end{figure*}

We  have also employed histograms to visualize the distribution of forecasts errors of the SOFS and the SOFS+RF, as shown in Figures from \ref{fig:error_p1} to \ref{fig:error_p5}. As can be seen, the use of the RF post-processor changes the error distribution. Ideally the post-processor would generate a narrow distribution centered around zero. The error distributions obtained by application of the RF post-processor have on general a mean closer to zero in most cases, or lower bias, specially for SSH forecasts when compared to the SOFS alone. A comparison between absolute biases with and without the post-processing for each station is shown in Table \ref{tab:bias}, where the bias is computed by calculating the average error obtained in the respective case. The only case where the bias increased by applying the post-processor is for SSH in station P3, the same one that saw an increase in RMSE after application of the post-processor.

\begin{table}
\begin{center}
\begin{tabular}{lrrrrrr}
\hline
 & \multicolumn{3}{|c|}{ \textbf{SOFS}} & \multicolumn{3}{|c}{ \textbf{SOFS+RF}} \\ \cline{2-7}
\textbf{Station}& \multicolumn{2}{|c|}{ {Current}} &  & \multicolumn{2}{|c|}{ {Current}} &  \\ \cline{2-3} \cline{5-6}
& \multicolumn{1}{|r|}{Speed} & \multicolumn{1}{|r|}{Dir.} &  \multicolumn{1}{|r|}{SSH} & \multicolumn{1}{|r|}{Speed} & \multicolumn{1}{|r|}{Dir} & SSH \\ 
& \multicolumn{1}{|r|}{(m/s)} & \multicolumn{1}{|r|}{(\textdegree)} &  \multicolumn{1}{|r|}{(m)} & \multicolumn{1}{|r|}{(m/s)} & \multicolumn{1}{|r|}{(\textdegree)} & (m) \\ \hline
P1 & 0.037 & 19.2 & - & 0.001 & 16.4 & -\\ 
P2 & 0.030 & 22.7 &0.010 & 0.022 & 18.9 & 0.005 \\ 
P3 & 0.050 & 23.6 & 0.050 & 0.042 & 10.6 & 0.097\\ 
P4 & 0.054 & 18.1 & 0.034 & 0.041 & 1.5 & 0.019 \\ 
P5 & 0.027 & 5.3 & 0.029 & 0.006 & 3.4 & 0.004\\ \hline
\end{tabular}

\end{center}
\caption{Comparison of the error distribution absolute bias for the SOFS and the SOFS+RF.} \label{tab:bias}
\end{table}

The improvement due to the post-processor is most evident in the histograms in cases where the SOFS model has does not produce accurate forecasts, such as the case for current directions in P1 and P5, where the lower current speeds make the direction forecast more prone to error, or in P4, where the SOFS model has a positive or negative error depending if the current is entering or exiting the channel, producing a two peaked distribution. When compared to previous results in the region, the decrease in bias obtained from this post processor is smaller, but this might be due to the fact that the train and test datasets were split into two continuous time-series in this paper, while they were split at random in the previous study.

\section{Conclusions}
\label{sec:conclusions}

In this paper we have presented a Physics-Informed Machine Learning approach to reduce the forecasting error of surface currents and SSH in the Santos Channel, Brazil, by predicting the forecasting {\em error} of the SOFS model, a numerical model already in use for the region. 

In our PIML approach we used Random Forests (RF) to predict the SOFS error by taking into account, as inputs, both the SOFS forecasts and sensor data. An RF model was trained for each combination of target variable (current speed, current direction and SSH) and channel location (points P1 to P5 in Figure \ref{fig:map}), resulting in 15 different models. 
The best hyperparameters for each RF were found by 5-fold cross-validation in a training dataset containing 80\% of the data, using the Index of Agreement (IOA) as a metric for performance. Once the RFs are trained, they can be used as a post-processor to correct the SOFS forecasts.

%The best hyperparameters for each RF model were found using random search in a 5-fold Cross-Validation approach on a training dataset containing 80\% of the data, and using the Index of Agreement (IOA) as a performance metric. Once the RFs are trained, they were used as a post-processor to correct the predictions of the SOFS model.

This post-processor was tested on a dataset not previously seen in training, and its performance was measured using IOA and RMSE as metrics. 
The results show that the use of that post-processor increased IOA in all stations for SSH, in 4 out of 5 for current direction, and in 2 out of 5 stations for current speed, while the RMSE decreased for all combinations of target variables and stations, except in the case of the SSH for the P3 station. One possibility is that the increase in RMSE for that specific case is due to distribution shift, but further analysis is still required to find the root causes.

Several opportunities arise as this investigation continues. As an immediate next step, we intend to improve the proposed PIML architecture with tests and comparisons using other ML models as post-processors.
We also intend to experiment with architectural modifications so that more information can be used as inputs to the post-processor, as in this version our post-processor uses measured data obtained only immediately before the first prediction step. 
Another proposal is to use Long Short-Term Memory (LSTM) Neural Networks or Transformers to encode an arbitrary long time series of measurements taken just before the prediction event into a fixed-length input that will provide more information for predictions.

\section*{Acknowledgements}

% This work was carried out at the Center for Artificial Intelligence (C4AI-USP), with support by the São Paulo Research Foundation (FAPESP) under grant number 2019/07665-4 and by the IBM Corporation. This work is also supported in part by FAPESP under grant number 2020/16746-5, the Brazilian National Council for Scientific and Technological Development (CNPq) under grant numbers 310127/2020-3, 312180/2018-7, 310085/2020-9 and Coordination for the Improvement of Higher Education Personnel (CAPES). The authors also thank the Santos Marine Pilots for providing crucial data for this research. This work was also financed in part by the \textit{Coordenação de Aperfeiçoamento de Pessoal de Nível Superior} (CAPES Finance Code 001), Brazil, and by \textit{Ita\'{u} Unibanco S.A.} through the \textit{Programa de Bolsas Ita\'{u}} (PBI) program of the \textit{Centro de Ci\^{e}ncia de Dados} (C$^2$D) of \textit{Escola Polit\'{e}cnica} of USP. We also gratefully acknowledge partial support from CNPq (grants ). %%Alunos da Anna -- Do Cozman também tem que por CNPq e ver o num dele

This work was carried out at the Center for Artificial Intelligence (C4AI-USP), with support by the São Paulo Research Foundation (FAPESP grant number 2019/07665-4) and by the IBM Corporation. This work is also supported in part by FAPESP (grant number 2020/16746-5), the Brazilian National Council for Scientific and Technological Development (CNPq grant numbers 310085/2020-9, 310127/2020-3, and 312180/2018-7), Coordination for the Improvement of Higher Education Personnel (CAPES, Finance Code 001), and by \textit{Ita\'{u} Unibanco S.A.} through the \textit{Programa de Bolsas Ita\'{u}} (PBI) program of the \textit{Centro de Ci\^{e}ncia de Dados} (C$^2$D) of \textit{Escola Polit\'{e}cnica} of USP.

\
%% The file named.bst is a bibliography style file for BibTeX 0.99c
\bibliographystyle{named}
\bibliography{references}

\end{document}